\RequirePackage{lineno}
\documentclass[12pt]{iopart}

\usepackage{iopams}
\usepackage{graphicx}

\begin{document}

\title[Characteristics of Parton Energy Loss Studied with
High-$p_\mathrm{T}$ Particle Spectra]{\boldmath
  Characteristics of Parton Energy Loss Studied with
  High-$p_\mathrm{T}$ Particle Spectra from PHENIX \unboldmath}

\author{Klaus Reygers for the PHENIX\footnote{For the full list of
    PHENIX authors, see appendix 'Collaborations' of this volume.}
  collaboration}

\address{University of M{\"u}nster, Institut f{\"u}r Kernphysik,
Wilhelm-Klemm-Stra{\ss}e 9,\\ 48149 M{\"u}nster, Germany}
\ead{reygers@uni-muenster.de}
\begin{abstract}
  %\linenumbers 
  In the first three years of the physics program at the
  Relativistic Heavy Ion Collider (RHIC) a picture was established in
  which the suppression of hadrons at high transverse momenta
  ($p_\mathrm{T}$) in central Au+Au collisions is explained by energy
  loss of quark and gluon jets in a medium of high color-charge
  density.  Measurements of single particle spectra for a smaller
  nucleus (Cu), for different center-of-mass energies and with higher
  statistics were performed in the subsequent years and are used to
  test predictions and assumptions of jet quenching models in more
  detail.  The measurements presented here are consistent with a
  parton energy loss scenario so that these models can be used to
  relate the observed suppression to properties of the created medium.
\end{abstract}

%\maketitle

%\linenumbers
\section{Introduction}
The study of p+p, d+Au, and Au+Au collisions in the first three years
at RHIC showed that (i) hadrons at high $p_\mathrm{T}$ ($p_\mathrm{T}
\gtrsim 6$\,GeV/$c$) produced in central Au+Au collisions are strongly
suppressed, (ii) direct photons in Au+Au collisions at high
$p_\mathrm{T}$ are not suppressed, and (iii) high-$p_\mathrm{T}$
hadrons in d+Au are not suppressed \cite{Adcox:2004mh}. The particle
production in collisions of two heavy-ion species A+B with respect to
p+p collisions was quantified with the nuclear modification factor
$R_\mathrm{AB} =
(\left. \mathrm{d}N/\mathrm{d}p_\mathrm{T}\right|_{A+B})/(\langle
T_\mathrm{AB} \rangle \times
\left. \mathrm{d}\sigma/\mathrm{d}p_\mathrm{T} \right|_{p+p}) $ where
the nuclear overlap function $T_\mathrm{AB}$ is related to the number
of independent nucleon-nucleon collisions according to $\langle
T_\mathrm{AB} \rangle = \langle N_\mathrm{coll} \rangle
/\sigma_\mathrm{inel}^\mathrm{NN}$.

These observations provided strong evidence for jet quenching: quarks
and gluons lose energy in a medium of high color-charge density
created in Au+Au collisions whereas direct photons leave the medium
unscathed and follow the scaling with $\langle T_\mathrm{AB} \rangle$
since they interact only electro-magnetically. With high-$p_\mathrm{T}$
single particle spectra measured after the initial three runs at RHIC
assumptions and predictions of parton energy loss models can be
studied in more detail. In particular, the centrality and $p_\mathrm{T}$
dependence of hadron and direct-photon production can be tested at
higher $p_\mathrm{T}$. Moreover, the new data allow to test the dependence of
the hadron suppression on the heavy-ion species (Cu instead of Au), on
the hadron species, and on $\sqrt{s_\mathrm{NN}}$.

\section{Results}
% pT dependence of pi0 and direct gamma R_AA in Au+Au (and Cu+Cu)
\begin{figure}[t]
\includegraphics[width=\textwidth]{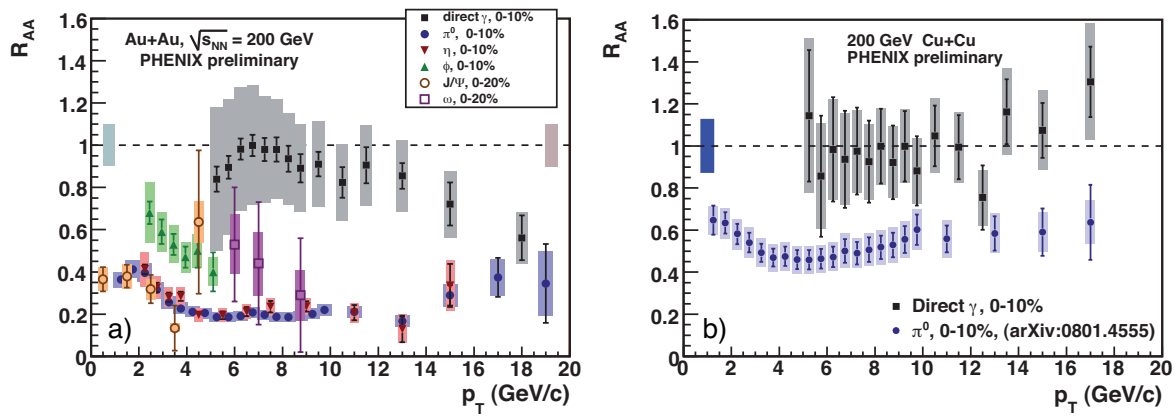}
\caption{a) $R_\mathrm{AA}$ in central Au+Au collisions at
  $\sqrt{s_\mathrm{NN}} = 200$\,GeV for direct photons and various
  mesons ($\pi^0$, $\eta$, $\phi$, $J/\Psi$, $\omega$).  b)
  $R_\mathrm{AA}$ in central Cu+Cu collisions at $\sqrt{s_\mathrm{NN}}
  = 200$\,GeV for direct photons and $\pi^0$'s.}
\label{fig:raa}
\end{figure}
Neutral-pion spectra in p+p and central Au+Au collisions at
$\sqrt{s_\mathrm{NN}} = 200$\,GeV, measured up to $p_\mathrm{T}
\approx 13$\,GeV/$c$ in the second physics run at RHIC, are now
available up to $p_\mathrm{T} \approx
20$\,GeV/$c$. Fig.~\ref{fig:raa}a shows that the observed suppression
remains approximately constant at $R_\mathrm{AA} \approx 0.2$ up to
highest $p_\mathrm{T}$ \cite{Adare:2008qa}.  Direct-photon yields in
central Au+Au collisions at $\sqrt{s_\mathrm{NN}} = 200$\,GeV, now
available up to $p_\mathrm{T} \approx 18$\,GeV/$c$, do not appear to
scale with $T_\mathrm{AB}$ at the highest $p_\mathrm{T}$
($R_\mathrm{AA} \approx 0.6$). Possible explanations of this
observation include the difference between the parton distributions in
protons and neutrons (isospin effect), the modification of the parton
distributions in nuclei (EMC effect), and the suppression of direct
photons which result from the fragmentation of partons.
Interestingly, a suppression of direct-photon production at
$p_\mathrm{T} \approx 17$\,GeV/$c$ is not observed in central Cu+Cu
collisions at the same energy (Fig.~\ref{fig:raa}b).

% dependence on particle species
The $R_\mathrm{AA}$ for different mesons in Fig.~\ref{fig:raa}a shows
that not all mesons are suppressed by the same factor.  Neutral pions
and $\eta$'s exhibit the same suppression which is consistent with a
picture in which these particles are produced in the fragmentation of
partons outside the hot and dense medium. The amount of suppression
for $J/\Psi$'s at mid-rapidity is similar to that of $\pi^0$'s and
$\eta$'s. However, $\omega$ and $\phi$ mesons appear to be less
suppressed. This interesting pattern provides an important test for
jet quenching models.
  
\begin{figure}[t]
\includegraphics[width=\textwidth]{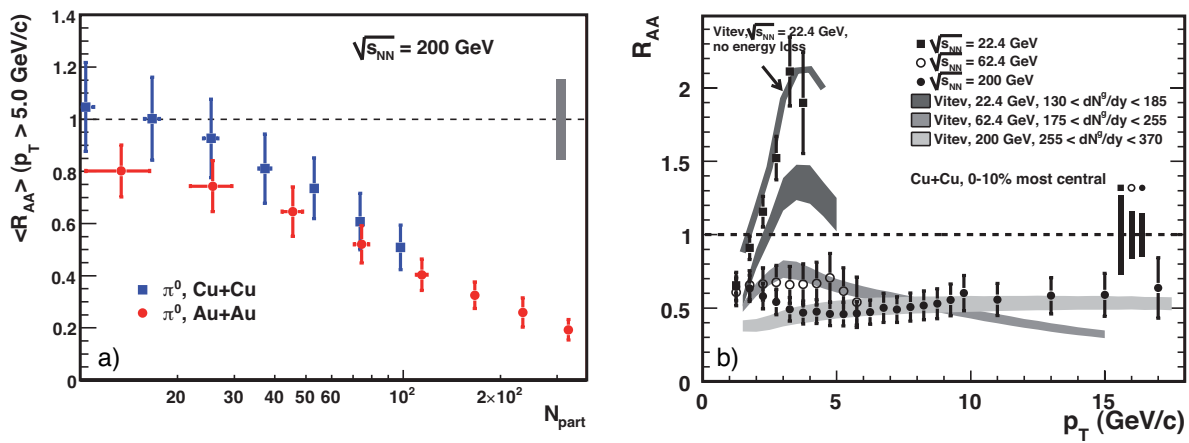}
\caption{a) Centrality dependence of the $\pi^0$ $R_\mathrm{AA}$ in
  Cu+Cu and Au+Au at $\sqrt{s_\mathrm{NN}} = 200$\,GeV. b) Energy
  dependence of the $\pi^0$ $R_\mathrm{AA}$ in central Cu+Cu
  collisions. The shaded bands represent parton energy loss
  calculations \cite{Adare:2008cx}.}
\label{fig:edep}
\end{figure}

% dependence on heavy ion species and centrality
The comparison of the $\pi^0$ suppression in Au+Au and Cu+Cu
collisions at $\sqrt{s_\mathrm{NN}}$ unveils a simple scaling: The
suppression only depends on the number of participating nucleons
($N_\mathrm{part}$) for the same $\sqrt{s_\mathrm{NN}}$ as shown in
Fig.~\ref{fig:edep}a. Such a scaling with $N_\mathrm{part}$ is
consistent with a parton energy loss picture \cite{Vitev:2005he}.
Fitting the centrality dependence of $R_\mathrm{AA}$ in central Au+Au
collisions for $p_\mathrm{T} > 10$\,GeV/$c$ with the function $R_{AA}
= (1 - \kappa N_\mathrm{part}^{\alpha})^{n-2}$ yields $\alpha = 0.56
\pm 0.10$, consistent with $\alpha \approx 2/3$ expected in parton
energy loss scenarios \cite{Adare:2008qa, Vitev:2005he}.

% energy dependence
The $\sqrt{s_\mathrm{NN}}$ dependence of the $\pi^0$ suppression
provides further constraints for parton energy loss models. A
comparison of the $\pi^0$ production in central Au+Au collisions at
62.4 and 200\,GeV shows that for $p_\mathrm{T} \gtrsim 6$\,GeV/$c$ the
$R_\mathrm{AA}$ at 62.4\,GeV approaches the $R_\mathrm{AA}$ at 200\,GeV
\cite{KleinBosing:2007bp}. This suggests that the smaller parton
energy loss at 62.4\,GeV is offset by the steeper parton $p_\mathrm{T}$
spectrum at this energy. The measurement of $R_\mathrm{AA}$ in Cu+Cu
at 22.4, 62.4, and 200\,GeV (Fig.~\ref{fig:edep}b) indicates that
parton energy loss starts to dominate over Cronin enhancement between
22.4\,GeV and 62.4\,GeV \cite{Adare:2008cx}. The $\pi^0$ suppression
at 62.4 and 200\,GeV is consistent with a parton energy loss
calculation in which the initial gluon density is derived from
measured charged particle multiplicities \cite{Vitev:2005he}. The
enhancement at 22.4\,GeV is consistent with a scenario without parton
energy loss.

\begin{figure}[t]
\includegraphics[width=\textwidth]{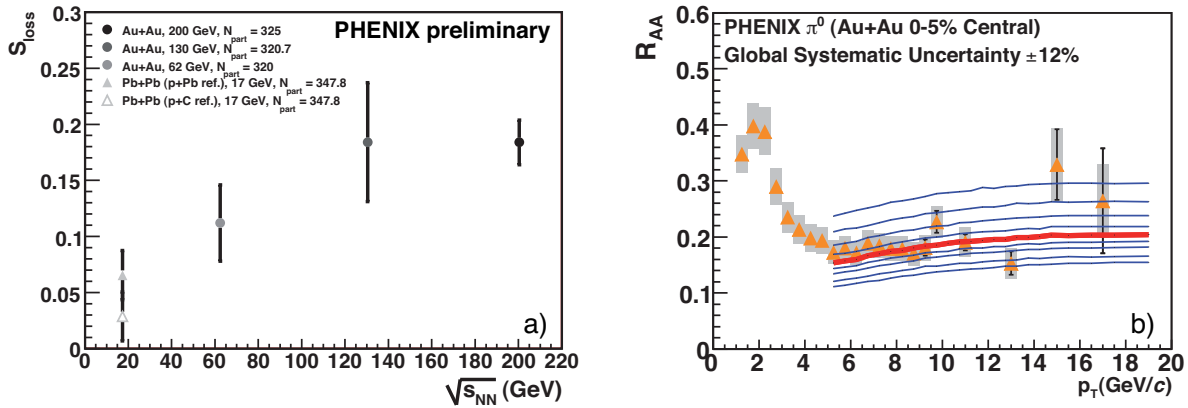}
\caption{ a) Energy dependence of $S_\mathrm{loss}$ for central Au+Au
  (RHIC) and Pb+Pb (CERN SPS) collisions as determined from the
  average $R_\mathrm{AA}$ in the range $2.5 \lesssim p_\mathrm{T}
  \lesssim 3.5$\,GeV/$c$.  b) Parton energy loss calculation
  \cite{Zhang:2007ja} for various values of the initial energy density
  $\varepsilon_0$ compared to the measured $\pi^0$ $R_\mathrm{AA}$ at
  $\sqrt{s_\mathrm{NN}} = 200$\,GeV.}
\label{fig:sloss}
\end{figure}

% sloss
Characterizing the parton energy loss $\Delta E$ with $R_\mathrm{AA}$
is problematic since $R_\mathrm{AA}$ not only depends on $\Delta E$
but also on the steepness of the parton $p_\mathrm{T}$ spectrum.  The
parton $p_\mathrm{T}$ spectra become steeper with decreasing
$\sqrt{s_\mathrm{NN}}$ leading to a smaller $R_\mathrm{AA}$ for the
same $\Delta E$. This might explain why a significant $\pi^0$
suppression ($R_\mathrm{AA} \approx 0.5 - 0.6$) is even observed at
$\sqrt{s_\mathrm{NN}} = 17.3$\,GeV in very central Pb+Pb collisions
\cite{Aggarwal:2007gw}. For a power law spectrum
$1/p_\mathrm{T}\,\mathrm{d}N/\mathrm{d}p_\mathrm{T} \propto
p_\mathrm{T}^{-n}$ and an approximately constant $R_\mathrm{AA}$ the
fractional parton energy $S_\mathrm{loss} = \Delta
p_\mathrm{T}/p_\mathrm{T}$ loss can be estimated as $S_\mathrm{loss} =
1 - R_\mathrm{AA}^{1/(n-2)}$ where $n$ is obtained from a power law
fit of the p+p spectrum \cite{Adler:2006bw} at the same energy.  At
$\sqrt{s_\mathrm{NN}} = 17.3$, 62.4, 130 and 200\,GeV the values 11.4,
9.2, 8.2, 8.2, respectively, were used for the power $n$. For the CERN
SPS energy of 17.3\,GeV a $\pi^0$ spectrum measured in p+C collisions
was used as a replacement for a p+p spectrum
\cite{Aggarwal:2007gw}. As a function of $\sqrt{s_\mathrm{NN}}$
$S_\mathrm{loss}$ appears to increase smoothly with
$\sqrt{s_\mathrm{NN}}$ in central collisions with $N_\mathrm{part}
\gtrsim 320$ and reaches $\sim 0.2$ at the top RHIC energy of
$\sqrt{s_\mathrm{NN}} = 200$\,GeV (Fig.~\ref{fig:sloss}a).

% constraints for medium parameters
In order to constrain medium properties with the aid of parton energy
loss models a fitting procedure was described in \cite{Adare:2008cg}
that takes different kinds of systematic uncertainties
(point-by-point uncorrelated, point-by-point correlated, and overall
normalization uncertainties) into account. As an example, results of
the calculation described in \cite{Zhang:2007ja} are shown in
Fig.~\ref{fig:sloss}b for different values of the energy loss
parameter $\varepsilon_0$. The optimal fit (thick line) corresponds to
$\varepsilon_0 = 1.9^{+0.2}_{-0.5}$\,GeV/fm (one standard
deviation). In general, for the models described in
\cite{Adare:2008cg} medium parameters are constrained within
$20-25\,\%$ at the $1\,\sigma$ level. However, it needs to be stressed
that at this stage the theoretical uncertainties in the models are
much larger so that, {\it e.g.}, the values obtained for the medium
parameter $\hat{q}$ can differ by an order of magnitude.
  
\section{Conclusions}
The flat $R_\mathrm{AA}(p_\mathrm{T})$ up to $p_\mathrm{T} \approx
20$\,GeV/$c$ and the centrality dependence in Au+Au collisions at
$\sqrt{s_\mathrm{NN}} = 200$\,GeV, the scaling of $R_\mathrm{AA}$ with
$N_\mathrm{part}$ in Cu+Cu and Au+Au for the same
$\sqrt{s_\mathrm{NN}}$, the identical suppression for $\pi^0$'s and
$\eta$'s, and the energy dependence of the $\pi^0$ production in Cu+Cu
are observations which are all consistent with a parton energy loss
picture. More work, however, is needed to understand the different
suppression patterns of $\phi$ and $\omega$ mesons as compared to
pions and $\eta$'s and the scaling of direct photons at high
$p_\mathrm{T}$ in central Cu+Cu and Au+Au collisions.

\end{document}